\documentclass[pra,twocolumn,a4paper,showpacs,floatfix]{revtex4}
\usepackage{graphicx}
\begin{document}
\title{Electronic transmission in a model quantum wire with side coupled quasiperiodic chains: Fano resonance 
and related issues} 
\author{Arunava Chakrabarti}  
 \affiliation{Department of Physics \\University of Kalyani \\ 
Kalyani, West Bengal 741 235, India.}
\begin{abstract}
We present exact results for the transmission coefficient of a linear lattice at one or more sites 
of which we attach a Fibonacci quasiperiodic chain.
Two cases have been discussed, viz, when a single quasiperiodic chain is coupled to a 
site of the host lattice and, when more than one dangling chains are grafted periodically
along the backbone. 
Our interest is to
observe the effect of increasing the size of the attached quasiperiodic chain on the 
transmission profile of the model wire. We find clear signature that, 
with a side coupled semi-infinite Fibonacci chain, the Cantor set
structure of its energy spectrum should generate interesting multifractal character in the
transmission spectrum of the host lattice. This gives us an opportunity to control the 
conductance of such systems and to devise novel switching mechanism that can act over arbitrarily small
scales of energy. The Fano profiles in resonance are observed at various intervals of energy as well.
Moreover, an increase in the number of such dangling chains may lead to the design of a kind 
of spin filters. This aspect is discussed.
\end{abstract}
\maketitle 
\section{Introduction}
Electronic properties of simple one and quasi-one dimensional model systems have been studied
with keen interest over the past few years. This is mainly due to the success of nanofabrication
techniques in producing extremely small quantum objects characterised by discrete energy levels and 
novel transport properties, which can be
examined in a controllable way \cite{gold98}-\cite{shang01}. Precision instruments
such as the scanning tunnel microscope (STM) can now be used to build low dimensional 
nanostructures with tailor made geometries, such as one dimensional wires and 
confined two dimensional structures \cite{sund91}-\cite{gam00}. The tip of an STM 
can act as tweezers and can practically `place' individual atoms on a substrate. Thus, 
simple low dimensional model quantum systems
 are now closer to reality. 

In recent times, there has been some attention on the study of electronic transport
properties of linear atomic chains with side coupled finite segments in the forms
of chains or rings \cite{vas98}-\cite{dom04}. Motivated by the ability of an STM 
to place metal atoms along a line on an insulating substrate, thus creating what one may call an ideal 
metallic chain,  
Vasseur et al \cite{vas98} have theoretically studied the spectral properties of 
periodic arrays of atomic sites attached to linear chains to 
simulate the band structure of one dimensional atomic wires of alkali elements. 
Pouthier and 
Girardet \cite{pou02} have reported transport studies on similar systems, but with the side 
attached chains occupying randomly selected positions along the quantum wire. The resonance 
and anti-resonance have been studied in details on the basis of a scaling theory. 
Within the tight binding formalism, a model of a quantum wire
with a side coupled array of non-interacting quantum dots has been persued by 
Orellana et al \cite{ore03} and later, by Dominguez-Adame et al \cite{dom04}. The conductance 
oscillations with an odd-even parity effect, depending on the number of `quantum dots' 
in the dangling chain have been observed. Interestingly, similar models 
involving a random distribution of finite sized chains attached to different 
sites of a lattice (the backbone) were studied earlier
by Guniea and Verg\'{e}s \cite{gun87}. 
They discussed the localization properties exhibited by such systems.

While the localization, conductance and band structure of an ordered array of atomic sites with
one or more finite chains attached to it have exhibited non-trivial behavior, 
the observation of Fano resonance \cite{fano61} in the transmission properties of such systems 
has been another interesting aspect. The Fano lineshape in known to occur in different branches of physics, 
such as atomic physics \cite{fano61}, atomic photoionization \cite{fano81}, optical absorption
\cite{faist97}, Raman scattering \cite{cer73}, scanning tunneling through a surface impurity
atom \cite{mad98}. The manifestation of Fano resonance is in the form of
 a sharp asymmetric profile in transmission or absorption lines, arising out of 
the interference effect between the discrete energy level (caused by an impurity) and 
the continuum of states. 
The general formula describing the Fano lineshape is given by, 
\begin{equation}
\ {\cal F} (\omega) = \frac{(\omega+\delta)^2}{\omega^2 + 1}
\end{equation}
where, $\omega=(E-E_R)/(\Gamma/2)$. $E$ is the energy of the electron, 
 $E_R$ is the resonance energy and $\Gamma$ is the 
line width. $\delta$ is the so called `Fano factor' which controls the asymmetry in the line shape.

Very recently, it has been reported that similar Fano line shapes are observed in simple 
geometries such as a finite chain of atoms connected to one or more sites in a perfectly
ordered array of atomic sites. Using the Fano-Anderson model \cite{gdm93}, Miroshnichenko et al 
\cite{mir05} have studied the linear and non-linear Fano resonance in the transmission across a 
periodic array of identical sites at one point of which a single defect site is side coupled.
It is shown that the Fano resonance is observed as a specific feature in the transmission coefficient
 as a function of the frequency \cite{mir05}. Extending the idea to the case when the attached 
Fano defect is no longer a single atom, but a chain of atoms, Miroshnichenko and Kivshar 
\cite{kiv05} have 
proposed a possible engineering of Fano resonances. Such simple models, as the authors argue, 
can be used to model discrete networks of coupled defect modes in photonic crystals and 
complex waveguide arrays and ring-resonator structures. Earlier, Guevara et al \cite{lad03}
and Orellana et al \cite{ore04} had studied electronic transport through a model of a 
double quantum dot molecule attached asymmetrically to leads. The Fano and Dicke profiles
were observed and analysed, and the role of an applied magnetic field was discussed as well 
\cite{ore04}.

The above studies \cite{mir05}-\cite{ore04} are made within the framework of a tight binding 
model, the Hamiltonian being described in the Wannier basis. The `interaction' with the defect 
state is described by a hopping amplitude \cite{mir05}. Thus, in such works the coupling 
in energy space between the continuum of states and the discrete quasi-bound state, as in
Fano's original work \cite{fano61}, is alternatively viewed, in a real space description, as a `hopping'
between an atomic site in a perfect infinite lattice and that in a side attached chain, which 
we shall refer to as a `Fano defect'. What really
is interesting is that, such a simple real space description allows one to handle the problem
analytically, get some insight into how the Fano lineshapes are controlled by various parameters
of the system. These model studies thus, as observed by Miroshnichenko et al \cite{mir05}
`may serve as a guideline for the analysis of more complicated physical models'.

An important isuue however, has not been addressed so far, to the best of our knowledge. How 
does the nature of the distribution of eigenvalues of the isolated Fano defect 
affect the transmission through the quantum wire, when one or more Fano defects are coupled 
to the wire ? It is known that the zeroes of transmission occur at 
the discrete eigenvalues of the hanging chain when it is 
decoupled from the quantum wire \cite{dom04},\cite{mir05},\cite{kiv05}. Therefore, the spectral 
character of the isolated Fano defect is likely to produce interesting features in the transmission
spectrum. Motivated by this idea, in this communication, we investigate the transport across a 
tight binding chain of periodically placed atomic sites  
with a quasiperiodic defect chain side coupled to it. The defect chain in the present work is chosen to
be a Fibonacci chain \cite{kkt83}. In particular, we focus on the situation
when the length of the dangling Fibonacci chain extends to infinity. In this limit, the Fibonacci
chain, when isolated, will give rise to a singular continuous spectrum with the eigenvalues distributed
in a multifractal Cantor set \cite{kkt83}-\cite{macia00}. By properly choosing the parameters of the 
system the entire spectrum of the side coupled Fibonacci chain can be made to lie within the allowed 
band of the periodic lattice (our `quantum wire', in the spirit of Ref.\cite{dom04}). 
This may lead to interesting transmission properties.     
In what follows we demonstrate that indeed, very interesting transmission properties are obtained, 
in which continuous zones of high transmission are punctuated with sharp drops to zero transmission.
Such feature, for a long enough chain, is observed at various scales of energy, exhibiting a self-similar
pattern. The self similarity becomes pronounced at smaller and smaller intervals 
of energy as the size of the defect chain 
is increased. This might provide us with an option of designing molecular switches which can tune a 
system from an `on' to an `off' state over an arbitrarily small range of energy.

We have also investigated transport properties in presence of 
multiple hanging Fibonacci chains separated by an arbitrary number of sites. In particular, for two such 
hanging chains we have come across a special feature in transmittivity which may give rise to an idea
of designing spin filters with such geometries.
In what follows we describe our systems and the results.

In section II we describe the model. Section III provides the first results in the context of a 
single side coupled chain. In section IV we present the band structure and transmission when 
multiple chains are attached to the wire, and in section V we draw conclusions.
\section{The model}
We begin with a reference to Fig.1. The hanging chain with the shaded circles represent the
quasiperiodic Fano defect, which we will refer to as a Fibonacci-Fano (FF) defect. 
The Fibonacci chain is grown recursively by repeated application 
of the inflation rule \cite{kkt83}, $L \rightarrow LS$ and $S \rightarrow L$, where, $L$
and $S$ may stand for two `bonds'. The first few generations are, 
\begin{displaymath}
G_1 : L,   
G_2 : LS,  
G_3 : LSL, 
G_4 : LSLLS
\end{displaymath}
and, so on. At any $n$th genetarion, the total number of the `bonds' is a Fibonacci number $F_n$, where, 
\begin{equation}
F_n = F_{n-1} + F_{n-2}
\end{equation}
for $n \ge 2$ with $F_0=1$, and $F_1=1$.
Depending on the nearest neighbour topology, we identify  
three kinds of sites, viz, $\alpha$, $\beta$ and $\gamma$, flanked by $L-L$, $L-S$ and $S-L$
bonds. 
The Hamiltonian of the system, in the standard tight binding form, is written as, 
\begin{equation}
H = H_{Wire} + H_{Defect} + H_{Wire-Defect}
\end{equation}
where, 
\begin{eqnarray}
H_{Wire} & = & \epsilon_0 \sum_{i=-\infty}^{\infty} c_i^{\dag} c_i + t_0 \sum_{<ij>} c_i^{\dag} c_j \nonumber \\
H_{Defect} & = & \epsilon_T d_1^{\dag}d_1 + \sum_{i=2}^{N-1} \epsilon_i d_i^{\dag} d_i + \nonumber \\ 
& & \epsilon_B d_N^{\dag}d_N + \sum_{<ij>} t_{ij} d_i^{\dag} d_j \nonumber \\
H_{Wire-Defect} & = & t_c (c_0^{\dag}d_1 + d_1^{\dag}c_0) 
\end{eqnarray}
In the above, $c^{\dag}(c)$ and $d^{\dag}(d)$ represent the 
creation (annihilation) operators for the wire and the defect respectively.
$\epsilon_0$ and $t_0$ are the on-site potential and the constant hopping integral in the 
periodic array of sites, which is our `wire'. The on-site potential in the bulk of the defect Fibonacci
chain takes on three different values, viz $\epsilon_{\alpha}$, $\epsilon_{\beta}$ and 
$\epsilon_{\gamma}$ corresponding to the three kinds of sites referred to above. 
The first (top) and the last (bottom) sites of the hanging chain are assigned on-site 
potentials $\epsilon_T$ and $\epsilon_B$ respectively.
The nearest neighbor
hopping integral takes on two values $t_L$ and $t_S$ for electron hopping along an $L$ or an $S$ bond.
The Fibonacci chain is assumed to be connected to the site indexed `zero' in the quantum wire. 
$t_c$ represents the hopping integral between 
the zeroth site of the wire and the first site of the Fibonacci chain. So, $t_c$ is the `interaction'
which locally couples the two subsystems, viz, the periodic chain, and the dangling FF defect . 
If we decouple the two, the spectrum of the periodic quantum wire is
absolutely continuous, the band extending from $\epsilon_0-2t_0$ to $\epsilon_0+2t_0$. The dispersion
relation is $E=\epsilon_0+2t_0 \cos qa$, $a$ representing the lattice spacing. With non-zero $t_c$ it is 
not right to view the two subsystems separately. In this case, the defect chain may be looked upon as a 
single impurity with a complicated internal structure and located at a single site in an otherwise
periodic array of identical potentials $\epsilon_0$. Therefore, while examining the transmission across 
the impurity, one has to be careful to adjudge whether the concerned energy really belongs to the 
spectrum of the entire system, that is, the ordered backbone plus the FF defect.
\begin{center}
\begin{figure}
{\centering \resizebox* {8cm}{5cm}{\includegraphics{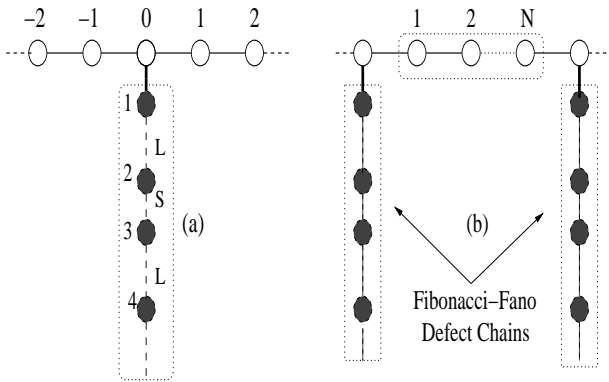}}}
\caption{\label{fig1}(a) The Fibonacci chain side coupled to a single site (b) The case
of multiple grafted chains. }
\end{figure}
\end{center}

To calculate electronic transmission across the quantum wire, we use the real space renormalization 
group (RSRG) method \cite{south83}. We renormalize the hanging Fibonacci chain by decimating the internal sites
selectively. For this we have to `fold' the chain backwards by using the growth rule in opposite
direction. To get unique recursion relation for the site at the chain's lower boundary, 
we consider Fibonacci chains with odd index of generation only. Thus, instead of the 
growth rule, as pointed out before, we fold the chain by using the rule $LSL \rightarrow L'$ and 
$LS \rightarrow S'$, which is one step ahead of the usual deflation rule $L \rightarrow LS$ and 
$S \rightarrow L$. This is to be appreciated that, it is not a limitation of the scheme. 
As our interest is
the large size behaviour of the impurity, by considering successive odd generations only of the 
Fibonacci chain we do not 
lose any physics whatsoever. Staring with a $2n+1$-th generation Fibonacci chain, decimation of $n$-steps
folds it to a diatomic molecule with two end atoms having on-site potentials $\epsilon_{T,n}$ and
$\epsilon_{B,n}$, and connected to each other by an effective hopping integral $t_{L,n}$. 
The process is outlined in the Appendix.
Using this method, we can create  
a single, {\it effective impurity} with site potential $\epsilon^*$ sitting at the $0$-th site of 
the ordered backbone (quantum wire). The recursion relations are given by, 
\begin{eqnarray}
\epsilon_{\alpha,n} & = & \epsilon_{\alpha,n-1} + \frac{A_{n-1}}{D_{n-1}} + \frac{B_{n-1}}{D_{n-1}} \nonumber \\
\epsilon_{\beta,n} & = & \epsilon_{\alpha,n-1} + \frac{t_{L,n-1}^2}{E_{\beta,n-1}} + 
\frac{B_{n-1}}{D_{n-1}} \nonumber \\ 
\epsilon_{\gamma,n} & = & \epsilon_{\gamma,n-1} + 
\frac{t_{S,n-1}^2}{E_{\beta,n-1}} +  
\frac{A_{n-1}}{D_{n-1}} \nonumber \\ 
\epsilon_{T,n} & = & \epsilon_{T,n-1} + 
\frac{A_{n-1}}{D_{n-1}} \nonumber \\ 
\epsilon_{B,n} & = & \epsilon_{B,n-1} + \frac{B_{n-1}}{D_{n-1}} \nonumber \\ 
t_{L,n} & = & \frac{t_{L,n-1}^2 t_{S,n-1}}{D_{n-1}} \nonumber \\
t_{S,n} & = & \frac{t_{L,n-1}t_{S,n-1}}{E_{\beta,n-1}}
\end{eqnarray}
where, 
\begin{eqnarray}
A_n & = & t_{L,n}^2 E_{\gamma,n} \nonumber \\
B_n & = & t_{L,n}^2 E_{\beta,n} \nonumber \\
D_n & = & E_{\beta,n}E_{\gamma,n}-t_{S,n}^2
\end{eqnarray}
Here, $E_j \equiv (E-\epsilon_j)$ for $j=\alpha$, $\beta$ or $\gamma$.
Finally, for the single impurity site we get, 
\begin{equation}
\epsilon^* = \epsilon_0 + \frac{t_c^2}{E-\tilde \epsilon}
\end{equation}
where, 
\begin{equation}
\tilde \epsilon = \epsilon_{T,n} + \frac{t_{L,n}^2}{E-\epsilon_{B,n}}
\end{equation}
The problem is now reduced to calculating transmission across a single $\delta$-like impurity with 
an effective potential $\epsilon^*$. This is easily done \cite{kiv05}. The transmission 
coefficient is given by, 
\begin{equation}
T = \frac{\alpha_q^2}{\alpha_q^2 + 1}
\end{equation}
where, $\alpha_q = [2t_0 (E-\tilde\epsilon) \sin (qa)]/t_c$.  
\begin{center}
\begin{figure}
{\centering \resizebox* {16cm}{12cm}{\includegraphics[angle=-90]{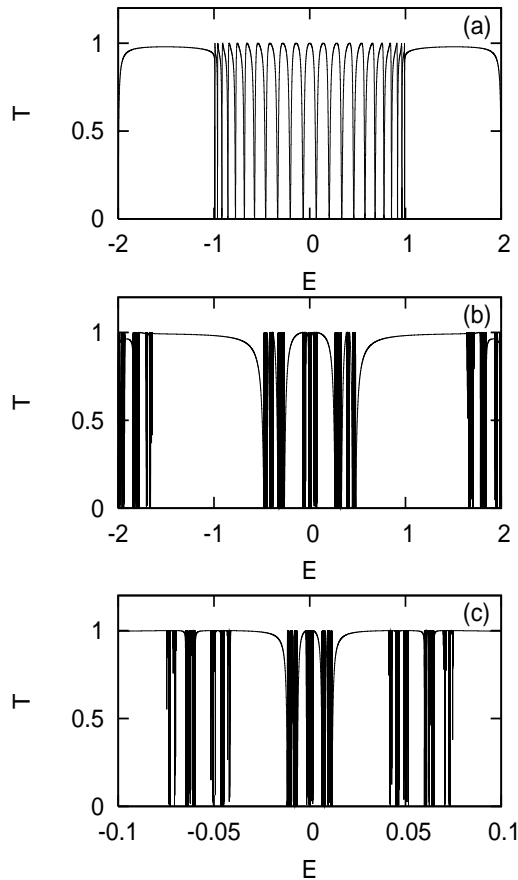}}}
\caption{\label{fig2}The transmission coefficient for (a) a periodic Fano defect with $21$ sites. The
hopping integral in the defect chain has been taken to be equal to $0.5$ in unit of $t_0$. 
(b) Transmission 
coefficient for a Fibonacci-Fano defect in the $23$rd generation. We have selected $\epsilon_j=0$ with
$j=\alpha$, $\beta$, $\gamma$. $t_L=t_S/2=0.5$ and $t_c=0.5$ in unit of $t_0$. 
The onsite potential $\epsilon_0$ and the hopping integral $t_0$ for the 
ordered quantum wire have been set equal to zero and one respectively.    
(c) The self similarity in the spectrum is highlighted.}
\end{figure}
\end{center}
Eq.(8) resembles the Fano formula (1), if we make the asymmetry parameter $\delta$ disppear.
The resonance line width is, 
\begin{equation}
\Gamma = \frac{t_c^2}{t_0 \sin (q_Ra)}
\end{equation}
with $q_R$ corresponding to the wave vector at resonance.
In this case, comparing with the Fano expression (1), we see that, 
$\omega=E-\tilde\epsilon$. $\tilde\epsilon$ is an energy dependent quantity. This, as we 
shall see, may give rise to asymmetric Fano profiles for a single hanging Fibonacci chain, 
as compared to a perfectly ordered chain.
\section{Results and discussion}
We begin by considering a single Fibonacci-Fano defect.
In Figs.2b and 2c we present results of the calculation of transmission coefficient across an effectively 
single defect that has been obtained after $n$-times renormalizing the Fibonacci chain of 
$2n+1$th generation. In particular, we discuss the `transfer model' of the Fibonacci chain with 
$\epsilon_\alpha = \epsilon_\beta = \epsilon_\gamma = 0$, and $t_L \ne t_S$. The energy spectrum of an 
infinite (or, semi-infinite) Fibonacci lattice with such specifications exhibits a three sub-band 
structure, each sub-band splitting up into three other sub-bands as one makes a closer scan in energy 
 \cite{kkt83}.
The result for an ordered chain is presented in Fig.2a for comparison.
Transmission zeroes are obtained from the roots of the equation 
\begin{equation}
E - \tilde\epsilon = 0
\end{equation}
At each real root of this equation $\epsilon^* \rightarrow \infty$, and a total reflection occurs.
For all real roots the incoming electron faces an infinitely high potential barrier. Naturally, 
among these roots we are concerned about those, which fall in the allowed band of 
the ordered backbone. The electron can not be transmitted at these energies, 
and we get transmission zeroes. 
It should be appreciated that these roots also yield the discrete eigenvalues of the isolated 
Fibonacci chain. In the thermodynamic limit the eigenvalues of the Fibonacci chain are 
distributed in a Cantor set of measure zero \cite{kkt83}. Therefore, when the length of the 
side coupled Fibonacci-Fano defect increases to infinity, we expect a Cantor set distribution 
of transmission zeroes in the $T$ vs $E$ diagram. 
The hierarchical 
structure of the spectrum is apparent in the transmission profiles, as shown in Fig.2c. 
By tuning the values 
of the parameters it is possible to include the entire spectrum of the isolated defect chain 
in the allowed band of the ordered chain playing the role of the quantum wire. The fragmented 
character of the spectrum of the defect Fibonacci chain then allows one to control the conductance  
of the wire with the defect, as every subcluster exhibits 
windows of finite transmission separated from each other by abrupt Fano drops to zero. 

Transmission resonances may be obtained from the solutions of the equation 
\begin{equation}
E - \epsilon_{B,n} = 0
\end{equation}
This makes $\tilde\epsilon=0$, and  $\epsilon^*=\epsilon_0$. Therefore, the `impurity' 
in this case becomes indistinguishable from a site in the perfectly ordered quantum wire.
For all such energy values the transmission coefficient $T=1$, independent of the wire-defect 
coupling $t_c$. The 
solutions of the equation $E-\epsilon_{B,n}$ are also distributed in a Cantor set structure. 

An essential difference in the transmission spectrum offered by a quantum wire with an attached 
ordered chain \cite{kiv05} and that with a single  attached Fibonacci-Fano defect is the observation of asymmetric
Fano profiles. To appreciate the difference we look at the expanded form of the effective 
impurity potential, viz, 
\begin{equation}
\epsilon_{0}^* = \epsilon_0 + \frac{t_{c}^2 [E-\epsilon_{B,n}]}{[E-\epsilon_{T,n}][E-\epsilon_{B,n}]
-t_{L,n}^2}
\end{equation}
As already noted, perfect transmission occurs at the energy eigenvalues satisfying
Eq.(12) 
and, the transmission zeroes are obtained from the solution of $E - \tilde\epsilon = 0$, or equivalently, 
\begin{equation}
[E - \epsilon_{T,n}][E - \epsilon_{B,n}] - t_{L,n}^2 = 0
\end{equation}
Owing to the highly fragmented nature of the spectrum of the 
isolated Fibonacci chain, any energy we hit upon will be arbitrarily close to a gap in the spectrum.
As a result, for a big enough Fibonacci-Fano defect, $t_{L,n}$ (and, $t_{S,n}$ as well), 
after $n$ steps of renormalization 
will become vanishingly small. Thus, if we deviate slightly from the resonance energy by setting 
$E=\epsilon_{B,n} + \Delta$, with $\Delta$ being an arbitrarily small number that we can choose at our will, 
the left hand side of Eq.(12) can be made arbitrarily close to zero. But, at the same time, $t_{L,n}$  
becomes very very small as well (for an appreciable size of the defect chain) , 
as the chosen energy, 
in all probability, will lie in a gap. 
Also, both $\epsilon_{T,n}$ and $\epsilon_{B,n}$ are of the same order of magnitude.
Thus, the left hand side of 
Eq.(15) can be made to have a magnitude at least $\sim \bigcirc (\Delta^2)$. 
This may make $\epsilon^* \rightarrow \infty$ and consequently, $T \rightarrow 0$ in the immediate 
neighborhood of a transmission resonance. This leads to the possibility of observing asymmetric 
profiles in the transmission spectrum with a Fibonacci-Fano defect chain.
For an ordered chain as a defect, 
all the energy values extracted from the equation $E-\epsilon_{B,n}=0$ correspond to extended eigenstates
of the chain (when isolated). As a consequence, 
the hopping integral under successive renormalization never flows to
zero for any such energy. In the thermodynamic limit, the spectrum of the isolated chain will be a 
continuum. That is, any deviation from the roots of the equation $E-\epsilon_{B,n}=0$ will also be 
in the spectrum, and would correspond to an extended state. The hopping integral remains non zero for
all such energies as well. This disapproves the juxtaposition of transmission zero and unit transmission, and 
hence an asymmetry in resonance profile.  
In Fig. 3 (bottom panel)  we highlight the concommitant zeroes which give rise to the asymmetry in lineshape in case 
of a Fibonacci-Fano defect in its $15$th generation ($987$ bonds). 
The uniformity in spacing of the eigenvalues of an ordered Fano defect 
does not allow for any asymmetry. This is shown in Fig.3 (top panel). 

The profile of the transmission window is sensitive to the choice of the wire-defect 
coupling $t_c$. For low $t_c$, we can have reasonably flat windows suddenly dropping down to 
provide zero transmission at $E = \tilde\epsilon$, while, 
with large enough values of $t_c$, 
the entire spectrum practically consists of sharp delta like transmission peaks arranged in a 
self similar triplet of clusters, typical of a Fibonacci lattice. The behavior is not difficult to 
understand if one remembers that $t_c$ is proportional to the width of resonance.

Before we end this section, it is worth mentioning that a non-zero flow of the hopping integrals 
, which usually implies the presence of extended eigenstates, may take place in a Fibonacci chain as well. 
One such energy, leading to a sharp asymmetric Fano 
resonance is shown in Fig.4. The defect chain is a $19$th generation Fibonacci lattice in the 
transfer model. Resonance takes place at $E=-0.03108$ in unit of $t_0$. The hopping integrals 
$t_{L,n}$ and $t_{S,n}$ do not flow to zero at this energy indicating that the corresponding 
eigenstate is not localized within the size of the defect. 
\begin{figure}
{\centering \resizebox* {12cm}{10cm}{\includegraphics[angle=-90]{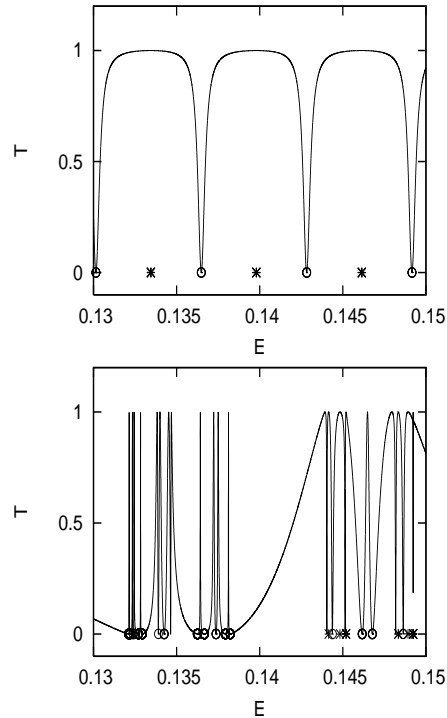}}\par}
\caption{\label{fig3}The transmittivity ($T$) 
against energy when a $15$th generation ordered chain (top panel)  and a  
Fibonacci chain (bottom panel) are side coupled. The positions of the resonance and 
anti-resonance are shown in each case by `asterix' and a circle with a dot.
The juxtaposition of the two symbols can be compared in the two cases.
All the parameters are as in Fig.2} 
\end{figure}
\section{Multiple side coupled Defects}
\subsection{Band Structure}

In the spirit of Vasseur et al \cite{vas98} and Pouthier and Girardet \cite{pou02} we 
have investigated the effect of attaching muliple Fibonacci-Fano (FF) defects of identical size at
regular interval on the ordered backbone. Just as in the case of an array of ordered side coupled 
chains \cite{vas98}, the FF defects, when grafted periodically along the backbone open up a 
series of absolute band gaps in the spectrum. For a Fibonacci chain in the $n$th generation we have 
$F_n$ bonds, and the band structure of an infinite array of grafted FF defects shows $F_n+N+1$ bands,
where $N$ in the number of sites in the ordered backbone which separate two neighboring FF defects.
\begin{figure}
{\centering \resizebox* {8cm}{5cm}{\includegraphics[angle=-90]{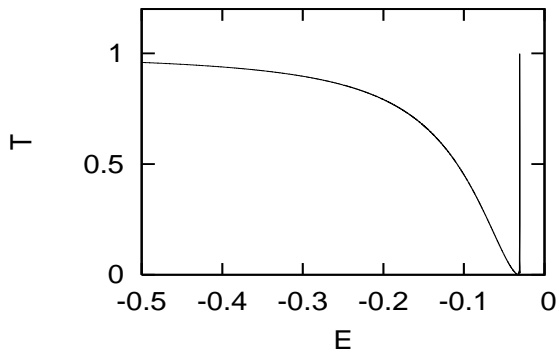}}}
\caption{\label{fig4}The sharply asymmetric Fano profile in the transmission spectrum
of a $19$th generation Fibonacci chain attached to the wire. 
The values of on-site potentials and the hopping integrals are as in Fig.2, and $t_c=0.5$ in 
unit of $t_0$.} 
\end{figure}
\subsection{Transmission Coefficient}

A simple algebra leads to the most general expression for the transmission coefficient 
of identical $n$ grafted chains separated from each other by $N$ atoms in the backbone. 
Each chain is renormalized into a single defect sitting on the backbone and separated from its nearest 
defect by $N$ `pure' sites. The system is further reduced, for convenience, 
to an array of identical pseudo-atoms by decimating the intermediate `pure' sites.
The new lattice spacing is now $(N+1)a$ and, the effective potential $\mu$ at each site as well as the 
effective nearest neighbour hopping integral $h$ are given by, 
\begin{eqnarray}
\mu & = & \epsilon^* + 2t_0 \frac{U_{N-1}(x)}{U_N(x)} \nonumber \\
h & = & \frac{t_0}{U_N(x)}
\end{eqnarray}
where, $x=(E-\epsilon_0)/2t_0$ and, $U_j$ is the $j$th order Chebyshev polynomial of the second kind.
It is now straightforward to compute the transmission coefficient of $n$ such defects embedded in a 
perfect lead.
The result is, 
\begin{widetext}
\begin{equation}
T = \frac{4 \sin^2 qa}{\left [-2U_{n+1}(y)+(rU_{n+2}(y)+\frac{U_{n}(y)}{r})\cos qa \right ]^2 +
\left [rU_{n+2}(y)-\frac{U_{n}(y)}{r}\right ]^2 \sin^2 qa}
\end{equation}
\end{widetext}
where, 
$r=h/t_0$, and, $y=(E-\mu)/2h$. 
In Fig.5 we show 
transmission spectrum for more that two Fibonacci-Fano defect chains. 
Two such cases are presented with $t_c=0.1$ and $1.0$ in unit of $t_0$. In each case there are
ten grafted chains separated by three sites of the backbone. Each grafted chain is a $15$th generation 
Fibonacci lattice containing $987$ bonds. With $t_c=0.1$, the absolute gaps are not formed with only $10$ 
chains. However, an increase in the value of $t_c$ to unity opens up gaps in the spectrum. With a 
small number of chains one may still look for asymmetric Fano profiles. However, quantum 
interference between the waves reflected back and forth between the grafted defect chains mask such 
resonances, and it is difficult to locate them numerically. The self similarity in the distribution 
of the eigenvalues of each defect, when they are decoupled from the backbone are lost as well.

\begin{figure}
{\centering \resizebox* {12cm}{10cm}{\includegraphics[angle=-90]{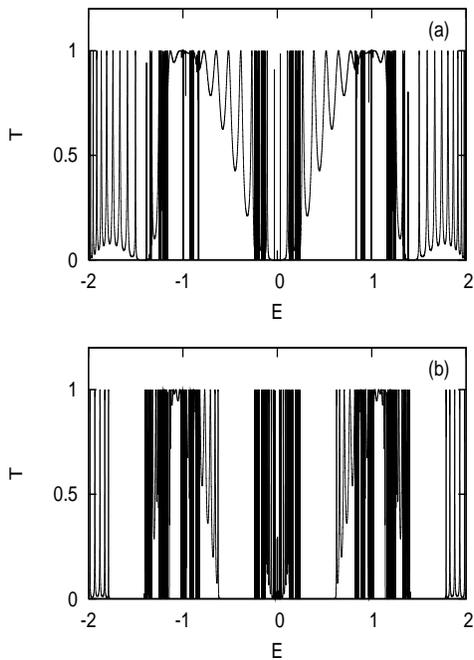}}}
\caption{\label{fig5}
Transmittivity (T) against energy (E) for ten $7$th generation Fibonacci-Fano defect chains
grafted in the ordered backbone. The chains are spaced $3$ `pure' sites apart. 
$t_c=0.1$ in (a), and $t_c=1.0$ in (b).
Other parameters 
of the system are as in Fig.2.} 
\end{figure}
\vskip .3in
\begin{center}
\it {Inversion of peak-dip positions: Two Fibonacci-Fano defects}
\end{center}
\vskip .25in
We now address separately the problem of transmission through two defect chains spaced $N$ sites apart.
It is found that, irrespective of
whether the defect chain is quasiperiodic or not, at special values of energy the positions of
resonance and the anti-resonance can be {\it swapped} by varying the number of `pure' atoms in the 
backbone separating the hanging defects. 
Incidentally, a similar phenomenon was also reported in  
Ref.\cite{kiv05} when an additional impurity atom was placed in the main array. Thus, it is 
sensible to attribute this feature (in systems such as we address here) to the presnece of 
two defects in an ordered backbone, at least one of which is a Fano defect. 
The significance of such swapping has already 
been pointed out elsewhere \cite{voochu}. For example, with a magnetic impurity, the transmission peak of the 
spin `up' (down) electrons may coincide with the transmission zero of the spin `down' (up) electrons.
This opens up the possibility of desiging a spin filter with such systems. It may be mentioned that, 
tight binding model of a quntum wire with an attached tight binding ring has already been proposed 
as a model of a spin filtering device \cite{lee06}.  

As an example of the above phenomenon, in Fig.6, we demonstrate the swapping of the peak-dip
positions for two cases. Fig.6a corresponds to the simpler one, where each of the two defect chains
consists of just two atoms coupled to the backbone at two different sites separated by $N$ intermediate
atomic sites of the backbone. The zeroes of transmission are at, $E= \epsilon_0 \pm t_0$.
The swapping has been demonstrated around $E=\epsilon_0 + t_0$ by changing $N$ from $N=3$ (solid line) to 
$N=4$ (dashed line). In Fig.6b, a similar transmission profile is exhibited for a Fibonacci-Fano defect 
with $21$ bonds. In each case, the swapping of the peak-dip profile takes place around a transmission-zero, 
which is obtained by solving the equation $E-\tilde\epsilon=0$.
Using a Fibonacci chain as a Fano defect opens up the possibility of getting  
a fractal distribution of such inversion of the resonance-anti resonance 
positions. One can perhaps then think of a spin filter that can operate on arbitrarily
narrow scales of energy. However, detecting such swapping of 
resonance profiles at various energies remains a non trivial task, 
as the spectra of the grafted chains, when quasiperiodic in character, offer a highly fragmented 
band structure. This aspect needs further attention.

Getting an anlytical insight into the phenomenon of swapping is difficult with big Fibonacci chains, as 
the expression for the transmission amplitude turns out to be a formidable one. 
However, we explicitly demonstrate that, 
for the case of the simpler diatomic chains, it is indeed possible to `see' the effect  
as $N$ changes.

Let us look back at Fig.1b with just two atoms in each of the defect chains.
We assign on-site potential $\epsilon_0$ to each atom in the two chains, while the inter-atomic hopping
in each chain has a value $t_0$. The expression for $\tilde \epsilon$, defined in Eq.(8) now becomes, 
\begin{equation}
\tilde \epsilon = \epsilon_0 + \frac{t_{0}^2}{E-\epsilon_0}
\end{equation}
With two such chains hanging $N$ atoms (of the backbone) apart, the problem is reduced to the 
calculation of electronic transmission through a system of $N+2$ atoms. 
The outer ones are the `impurity' atoms, each with site energy $\epsilon^*$ defined in Eq.(7). 
The $N$ `inner' atoms are `pure', and have on-site potential equal to $\epsilon_0$.
The nearest 
neighbour hopping integral is $t_0$ in this sample. The sample is then connected to two ideal semi-infinite
leads at the two ends, and the transmission amplitude can be worked out in a straightforward way \cite{stone81}
as,
\begin{equation}
\tau = \frac{e^{-ikNa} 2i \sin ka}{\alpha + i\beta}
\end{equation}
where, $\alpha = M_{11}+(M_{12}-M_{21}) \cos ka - M_{22} \cos 2ka$, $\beta = (M_{12}-M_{21}) \sin ka - 
M_{22} \sin 2ka$, and $M_{ij}$ are the elements of the transfer matrix \cite{stone81},
\begin{equation} 
M = M_{imp}.M_{pure}^N.M_{imp}
\end{equation}
Here, $M_{imp}$ and $M_{pure}$ stand for the transfer matrix across the `impurities' and
the `pure' site respectively. The explicit form of $M$ is, 
\begin{widetext}
\begin{equation}
M =  \left( \begin{array}{cc}(E-\epsilon^*)/t_0 & -1 \\1 & 0 \end{array} \right ).
  \left( \begin{array}{cc}U_N(x) & -U_{N-1}(x) \\U_{N-1}(x) & U_{N-2}(x) \end{array} \right ).
  \left( \begin{array}{cc}(E-\epsilon^*)/t_0 & -1 \\1 & 0 \end{array} \right )
\end{equation}
\end{widetext}
where, $U_N(x)$ stands for the $N$th order Chebyshev polynomial, already defined in Eq.(16).
We now look at the behavior of the transmission amplitude in the neighborhood of a transmission zero, 
which we choose to be $E=\epsilon_0+t_0$. We set $E=\epsilon_0+t_0+\sigma$ and neglect terms
$\sim \bigcirc (\sigma^2)$ or more (compared to $\sigma$ or terms free from $\sigma$) in the expanded 
matrix elements $M_{ij}$. The approximate expression of the transmission amplitude, in the neighborhood 
of the transmission zero becomes,
\begin{widetext}
\begin{equation}
\tau \sim \frac{4t_0i e^{-ikN} \sin k}{t_c^2 (U_N(x)-U_{N-1}(x))} \frac{\sigma^2}
{(\sigma - \frac{t_c^2}{2t_0} \frac{U_N(x)}{U_N(x)-U_{N-1}(x)}) 
+ i\sqrt{3}\frac{U_N(x)}{U_N(x)-2U_{N-1}(x)}\sigma}
\end{equation}
\end{widetext} 
The above expression immediately tells us that the `zero' of the numerator occurs at $\sigma=0$
($\sin k$ evaluated at the above energy is not going to be zero), while the zero of the real part 
of the denominator occurs at $\sigma=[t_c^2 U_N(x)]/[2t_0(U_N(x)-U_{N-1}(x)]=\delta$(say). The locations 
of the two zeroes are different. This set of detuned zeroes results in a non-zero Fano parameter and 
an asymmetric resonance around $E=\epsilon_0+t_0$. The quantity $\delta$ plays the role of the 
non-zero Fano parameter in this case. A swapping of the peak-dip positions can take place if the 
Fano parameter $\delta$ flips sign with a change in the value of $N$. We have chekcked that this 
precisely is the case here, as $N$ changes from three to four. In fact, different pairs of values of $N$ 
can be obtained for which the asymmetry parameter flips sign. 
As a result, when $\delta > 0$, the peak
appears after the dip, and when $\delta < 0$, the peak precedes the dip. 

It is worth mentioning that, the transition from the pole-zero (or, equivalently, peak-dip) to 
zero-pole positions was also investigated for model quantum systems by several authors \cite{kim99}
-\cite{vargia05}. The change in the asymmetry parameter was shown to be caused by a variation in 
some {\it continuous} parameter of the system, for example, the strength of the impurity in Ref.\cite{kim99}, 
or the magnetic flux in Ref.\cite{lad06}. On the other hand, in our case, 
the asymmetry parameter is a function of the discrete variable $N$, the number of 
sites between the two dangling chains. It is the change in the discrete
variable $N$ that causes the interesting swapping in the peak-dip positions. We have tested for other cases
also, namely, when each of the hanging chains contain three or four atoms. In each case, in a $\sigma$-
neighborhood of a transmission zero, we find $\tilde\epsilon \sim 1/\sigma$, leading to a $N$-dependent 
Fano parameter that flips sign when $N$ is made to change suitably. The case 
of two Fibonacci-Fano defects has also been examined numerically, and the result indicates 
that it is true for this system as well. Thus, we get confidence to 
predict that, the swapping is a generic feature of two Fano defects. Whether this remains true when more 
defect chains are incorporated, needs to be examined carefully. It may be mentioned that similar 
Fano-like profiles are also found in cases where the distribution of stubs follows a 
quasiperiodic sequence \cite{samar04}-\cite{sheelan05}. However, the aspect of Fano effects in transmission
was not addressed in any of these papers.

Before we end this section, a comment on the nature of the eigenstates will be in order. The 
most important thing to remember is that, to adjudge the character of an electronic state at a particular 
energy, one has to make sure that the energy concerned corresponds to an `allowed' state of the 
full system, i.e. the sample plus the semi-infinite leads. This implies that, 
a consistent solution of the Schr\"{o}dinger equation will exist at that energy all over the 
sample. It can be checked, at least for 
small Fano defects such as the case discussed above, by explicitly working out the amplitudes of 
the wave function at all sites of the system. It will be found that consistent solution can be obtained 
for various values of $N$, by assigning {\it zero} amplitude to the sites where the defect chains 
join the ordered backbone. The profile of amplitude will resemble a standing wave pattern between the
potential barriers (existing at the defect-backbone junction points) 
whose height in given by $\tilde\epsilon$. At $E=\tilde\epsilon$, the effective potential
$\epsilon^*$ becomes infinitely high. As a result, though a non-trivial distribution of amplitudes with 
finite values can be written throughout the system, the electronic transport reduces to zero. An electron
, released in any one of the intermediate $N$ sites and in between the
potential barriers, will be reflected back and forth. Beyond the well, however, one can have a perfectly 
periodic distribution of amplitudes. 
Thus, it is not 
right to call such states exponentially localized in the language of conventional localization studies.
As the nearest neighbouring sites are still coupled by non-zero 
hopping integral, it may be tempting to call them 
extended states. But, this will also be wrong, as the transmission across the sample for such an energy is zero. 
In a previous work, in the context of transport through a Vicsek fractal, such states were encountered, and have 
been termed `atypically extended' \cite{bb96}.

\begin{figure}
{\centering \resizebox* {12cm}{10cm}{\includegraphics[angle=-90]{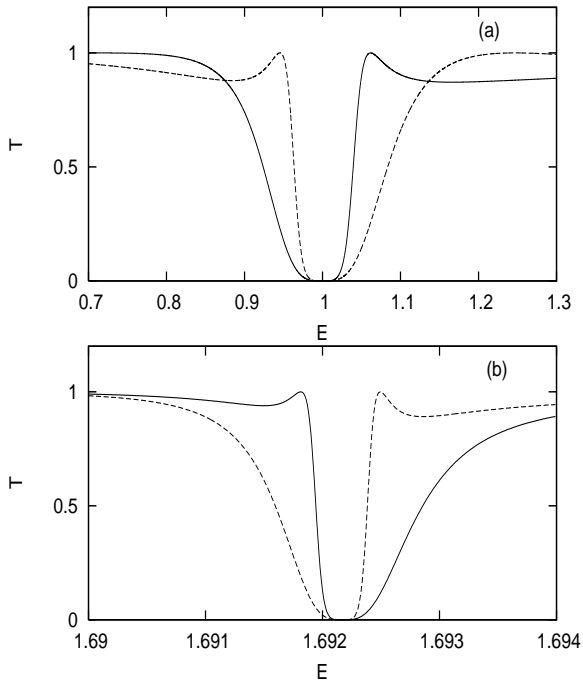}}}
\caption{\label{fig6}
Transmittivity of a two-defect sytem. (a) Two identical sites form a defect chain. Two such
chains are separated from each other by $N=3$ (solid line) and $N=4$ (dashed line) atoms.
(b) Two seventh generation 
Fibonacci-Fano defect chains ($21$ bonds) are attached to two sites of the quantum wire. 
The number of sites separating the two chains is $N=3$ (solid line)
and $N=6$ (dashed line). The parameters are, $\epsilon_0=0$, $t_0=1$, and, 
$\epsilon_\alpha=\epsilon_\beta=\epsilon_\gamma=0$, 
$t_L=1$, $t_S=2$ and $t_c=0.4$ in unit of $t_0$. The swapping of the resonance points are clearly seen.} 
\end{figure}
\section{Conclusion}
We have examined the effect of coupling a quasiperiodic chain of increasing length to an ordered
backbone on the electronic transmission. The self similar spectrum of the side coupled chain 
leads to an interesting transmission spectrum across the ordered lattice comprising of 
continuous zones of finite transmission inter-connected by sharp asymmetric Fano drops to zero.
This does not disregard the presence of Breit Wigner type of resonance profiles. But we have not 
discussed this. Within a framework of real space renormalization group method, we also discuss the 
band structure for an infinite array of FF defects. Finally the transmission properties when more than
one FF defect chains are attached, are discussed. With two chains there is a swapping of Fano resonance 
profiles is observed at a specific energy. This has been explained using a 
simpler system. This aspect, as also observed earlier, may be utilised in 
designing novel spin filters.

\begin{center}
{\bf Acknowledgement}
\end{center}
I thank Santanu Maity for useful suggestions in preparing the manuscript and Samar Chattopadhyay
for participating in some stimulating discussion.
\vskip .3in
\begin{center}
{\bf APPENDIX}
\end{center}
\vskip .25in
We briefly discuss the way to obtain Eq.(5) to Eq.(8). Real space decimation method 
(see \cite{sam02}, \cite{sheelan05}, and references therein) is used.
The basic set of difference equations to be used has the general shape, 
\begin{equation}
(E-\epsilon_j) \psi_j = t_{j,j-1} \psi_{j-1} + t_{j,j+1} \psi_{j+1}
\end{equation}
\begin{figure}
{\centering \resizebox* {7cm}{5cm}{\includegraphics{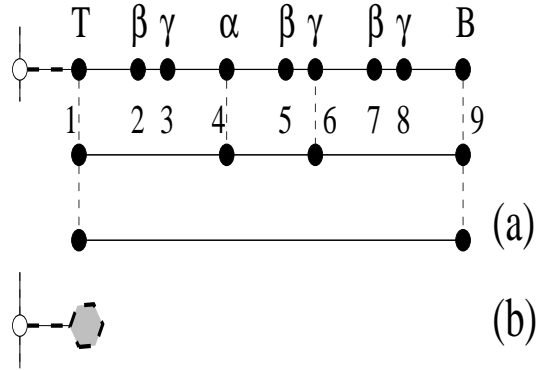}}}
\caption{\label{fig8}(a) Illustration of the basic decimation process. (b) Obtaining $\tilde\epsilon$ 
from the renormalized diatomic molecule.} 
\end{figure}

where, $\psi_j$ stands for the amplitude of the wave function at the $j$th site and $t_{j,j\pm 1}$
represents the nearest neighbour hopping integral between the $j$th and the $j \pm 1$th site.
As an illustration, we refer to Fig.8. $T$ stands for  `top' and $B$ stands for `bottom' atoms in 
Fig.1. A Fibonacci chain with eight bonds ($5$th generation) is attached to a site of the 
backbone (marked by the open circle in Fig.8a). The set of difference equations for the Fibonacci chain are, 
\begin{eqnarray}
(E-\epsilon_T) \psi_1 & = & t_c \psi_0 + t_L \psi_2 \nonumber \\
(E-\epsilon_\beta) \psi_2 & = & t_L \psi_1 + t_S \psi_3 \nonumber \\ 
(E-\epsilon_\gamma) \psi_3 & = & t_S \psi_2 + t_L \psi_4 \nonumber \\ 
(E-\epsilon_\alpha) \psi_4 & = & t_L \psi_3 + t_L \psi_2 \nonumber \\ 
(E-\epsilon_\beta) \psi_5 & = & t_S \psi_6 + t_L \psi_4 \nonumber \\ 
(E-\epsilon_\gamma) \psi_6 & = & t_S \psi_5 + t_L \psi_7 \nonumber \\ 
(E-\epsilon_\beta) \psi_7 & = & t_S \psi_8 + t_L \psi_6 \nonumber \\ 
(E-\epsilon_\beta) \psi_8 & = & t_L \psi_9 + t_S \psi_7 \nonumber \\ 
(E-\epsilon_B) \psi_9 & = & t_L \psi_8  
\end{eqnarray}
Here, $\psi_0$ implies the amplitude of wave function at the zeroth site of Fig.1 
(which is the open circle here) where, the defect
chain joins the back bone. The renormalization scheme $LSL \rightarrow L'$ and $LS \rightarrow S'$ 
implies that we eliminate the amplitudes at sites numbered $2 (\beta)$, $3 (\gamma)$, $5 (\beta)$, 
$7 (\beta)$ and $8 (\gamma)$ in terms of the remaining sites. This brings the $8$-bond FF defect to an 
effectively $3$-bond Fibonacci chain with renormalized values of the site energies and the hopping integrals.
Such a decimation scheme, when applied to a $2n+1$th generation Fibonacci chain, leads to the set of 
equations (5).
Consequently, the top and the bottom sites, which remain un-decimated, also have renormalized values of the 
on-site potential. One further decimation of the two intermediate sites (the second stage in Fig.8a) brings 
the entire chain down to a diatomic molecule with renormalized `top' and `bottom' atoms. Finally, the amplitude
at the bottom site is eliminated in terms of the amplitude at the top site to get the expression 
for $\tilde\epsilon$. This is depicted in Fig.8b.
With bigger FF defect chains the renormalization process is continued, till the final 
diatomic molecule (Fig.8a) is obtained. Thus we get $\tilde\epsilon$ in Eq.(8).

\end{document}